\documentclass[10pt,journal]{IEEEtran}
\usepackage{cite}
\usepackage{hyphenat}
\usepackage{graphicx}
\usepackage[T1]{fontenc}
\usepackage[latin1]{inputenc}
\usepackage{amsmath,amsfonts,amsbsy,amssymb}
\usepackage{mathabx}
\usepackage{mathrsfs}
\usepackage[nolist]{acronym}
\usepackage{float}
\usepackage{tipa}
\usepackage{tabularx}
\usepackage[colorlinks,linkcolor=black,bookmarksopen,bookmarksnumbered,citecolor=black,urlcolor=black]{hyperref}
\newtheorem{definition}{Definition}
\newtheorem{proposition}{Proposition}

\newtheorem{theorem}{Theorem}
\begin{document}
\newcommand{\erfc}[1]{
	\operatorname{erfc}\left(#1\right) 
}
\newcommand{\review}[1]{#1}
%
\title{On the Performance of Secure Full-Duplex Relaying under Composite Fading Channels} 
\author{%
Hirley Alves, 
Glauber Brante, 
Richard D. Souza, 
Daniel B. da Costa, 
and Matti Latva-aho, 
\thanks{H. Alves and M. Latva-aho are with the Centre for Wireless Communications, University of Oulu, Oulu, Finland. (emails: \{halves, matla\}@ee.oulu.fi).}
\thanks{H. Alves, G. Brante and R. D. Souza are with Federal University of Technology - Paraná (UTFPR), Paraná, Brazil (emails: \{gbrante, richard\}@utfpr.edu.br).}
\thanks{D. B. da Costa is with the Federal University of Ceará (UFC), Ceará, Brazil (email: danielbcosta@ieee.org).}
\thanks{This work was supported by the Finnish Funding Agency for Technology and Innovation (Tekes), University of Oulu Graduate School, Infotech Oulu Graduate School, Academy of Finland, CAPES/CNPq (Brazil).}
} 
%
\maketitle
%
\begin{abstract}
We assume a full-duplex (FD) cooperative network subject to hostile attacks and undergoing composite fading channels. We focus on two scenarios: \textit{a)} the transmitter has full CSI, for which we derive closed-form expressions for the \textit{average secrecy rate}; and \textit{b)} the transmitter only knows the CSI of the legitimate nodes, for which we obtain closed-form expressions for the \textit{secrecy outage probability}. We show that secure FD relaying is feasible, even under strong self-interference and in the presence of sophisticated multiple antenna eavesdropper. 
\end{abstract}
\section{Introduction}
Owing to the broadcast nature of the wireless medium, communication systems are  vulnerable to the attack of eavesdroppers and security issues play an important role in  system design. Recently, cooperative communications has appeared as an alternative to enhance physical layer (PHY) security \cite{LifengIT2008,DongTSP2010,AlvesICASSP2014}, where a trusted relay node helps the communication between Alice and Bob through cooperation or by jamming Eve. Most works focus on half-duplex (HD) schemes. However, full-duplex (FD) communications have gained considerable attention and several promising signal processing techniques for self-interference -- a power leakage between transmit and receive signals -- mitigation have been proposed \cite{DuarteFDTWC2012}. Even though ideal FD operation is not yet attainable, practical FD communication is feasible via such self-interference mitigation. This process results in residual self-interference that can be modeled as a fading channel, which allows the emulation of various (non) line-of-sight configurations \cite{DuarteFDTWC2012, AlvesWCL13}. 

In this work, we assume a FD cooperative network in the presence of a multi-antenna eavesdropper -- named Eve. Alice communicates with Bob -- which are the legitimate nodes -- with the help of a trusted FD relay, which suffers self-interference. We also employ a 
{general channel model which encompasses Nakagami-$m$ fading as well as Log-Normal shadowing,
which is known as composite fading channel \cite{ART:ILOW-TSP98}.} We consider two scenarios according to the availability of channel state information (CSI) at Alice: {\it a)} Alice has the CSI of all users, thus perfect secrecy can be achieved \cite{BlochBarros2008} and we focus on the {{average secrecy rate}} as the performance metric, for which we present an {{accurate approximation in closed-form}}; and {\it b)} Eve is a passive eavesdropper and Alice has only CSI of the legitimate channels. Thus, perfect secrecy cannot be guaranteed at all times \cite{BlochBarros2008}, and we provide {{accurate approximated closed-form expressions for the secrecy outage probability}}. To the best of the authors' knowledge, such analysis is still not available in the literature\footnote{In this work we assume a multi-antenna Eve, composite fading channels for two distinct scenarios, whereas \cite{AlvesICASSP2014} only presents a preliminary investigation on the secrecy outage probability of single antenna devices under Rayleigh fading channels.}. Additionally, besides the developed mathematical framework, we show that secure FD relaying is feasible in both scenarios under consideration, even under the effect of self-interference. 

\section{System Model}\label{sect:system_model}
We consider a cooperative network with three legitimate single-antenna users: Alice (A), relay (R) and Bob (B), communicating in the presence of Eve (E), which has $N_\text{E}$ antennas and employs maximal ratio combining (MRC). Moreover, we consider that the relay operates in FD mode (with one antenna dedicated to transmission and another to reception, in order to increase the isolation of the self-interference \cite{DuarteFDTWC2012}) and operates under the decode-and-forward (DF) protocol. 

The channels are affected by path-loss, shadowing and fading which are assumed to be  independent \cite{ART:ILOW-TSP98}. Then, in order to account for these channel impairments we adopt a composite fading distribution, where fading follows Nakagami-$m$ distribution and shadowing is modeled as Log-Normal (LN) random variable (RV), whose squared-envelop follows a Gamma-LN distribution \cite{ART:ILOW-TSP98}. Moreover, as proposed in \cite{ART:HO-ACM95}, the composite squared envelop -- which represents the SNR of a given link between nodes $i$ and $j$ -- is well approximated by a single LN RV, whose parameters depend on the actual distribution and are defined as: shape $\mu_\mathrm{dB} = \xi \left [ \psi\left ( m \right ) - \ln\left ( m \right)\right ] + \mu_{\Omega_p}$ and log-scale $\sigma_\mathrm{dB}^2 = \xi^2 \zeta\left ( 2, m \right) + \sigma_{\Omega_p}^2$, where $m$ is the shape parameter of the Nakagami-$m$ distribution, $\xi = \ln \left( 10 \right)/10$, $\Omega_p$ is the mean squared-envelop, $\mu_{\Omega_p}$  and $\sigma_{\Omega_p}$ is the mean and standard deviation of $\Omega_p$, respectively.
\subsection{Sum of LN RVs} As the density of $Z=\sum_{k} \gamma_{\text{ij}_k}$ -- a RV representing the sum of $k$ independent LN RVs -- has no exact closed form expression and its distribution is heavy-tailed and positively skewed, we resort to an approximation of the sum of LN RVs by a single LN RV \cite{PROC:WU-GLOBECOM05} as follows.

\begin{definition}[\textbf{Cumulants and additivity property}]\label{def:additivity}
\review{
Let $X$ and $Y$ be two independent RVs. The cumulants of a RV can be written as a function of the raw moments \cite{Petrov1975}. For instance, the first and second cumulants of $X$ are given as $\kappa_{1 , X} = \operatorname{E}\left[X \right]$ and $\kappa_{2 , X} = \operatorname{E}\left[X^2 \right]-\operatorname{E}\left[X \right]^2$ \cite{Petrov1975}. Then, the additivity property of the cumulants gives that the cumulants of $X+Y$ are the sum of the individual cumulants, therefore $\kappa_{n , X+Y} \triangleq \kappa_{n , X} + \kappa_{n , Y}$, where $\kappa_{n , X}$ and $\kappa_{n ,Y}$ are the $n$-th cumulants of $X$ and $Y$, respectively~\cite{Petrov1975}.
}
\end{definition}

\review{
From Definition~\ref{def:additivity} and by approximating the sum of LN RVs by a single LN RV \cite{PROC:WU-GLOBECOM05}, we estimate the parameters of the single LN RV from the cumulants as $\mu = \ln{\left(\kappa_1^2\right)}-\ln(\sqrt{\kappa_1^2 + \kappa_2})$ and $\sigma^2 = \ln{\left( \kappa_1^2 + \kappa_2 \right)}-\ln(\kappa_1^2)$, where $\mu$ is the mean and $\sigma^2$ is the variance of the equivalent $\mathsf{Normal}(\mu, \sigma^2)$ distribution in logarithmic scale \cite[Ch. 26]{ABRAM65,Petrov1975}.
}

\subsection{Legitimate Channel}
The DF protocol can be decomposed into two phases: broadcast (BC) and multiple access (MAC). Differently from HD relaying, the MAC phase starts simultaneously with the BC phase under the FD mode, so that the relay forwards to Bob the message received from Alice in a previous phase at the same time that Alice broadcasts a new message to the relay and Bob. Thus, the received signals at the relay and Bob are
\begin{align}
\label{eq:yAR}
{y}_{\text{R}} &= \sqrt{P_\text{A} d_{\text{AR}}^{-\nu}} \, h_{\text{AR}} \cdot { x} +\,\sqrt{ P_\text{R} \, \delta } \, h_{\text{RR}} \cdot { \tilde{x}} + { w}_{\text{R}},  \\
\label{eq:yAB}
{y}_{\text{B}} &= \sqrt{P_\text{A} d_{\text{AB}}^{-\nu}} \, h_{\text{AB}} \cdot { x} +\,\sqrt{ P_\text{R}\,d_{\text{RB}}^{-\nu} } \, h_{\text{RB}} \cdot { \tilde{x}} + { w}_{\text{B}},
\end{align}
where $h_{\text{ij}}$, $\text{i} \in \{\text{A} , \text{R} \}$ and $\text{j} \in \{\text{R} , \text{B}\}$, denotes channel coefficient, $P_\text{i}$ is the transmit power, $d_{\text{ij}}$ is the distance between nodes $\text{i}$ and $\text{j}$, and $\nu$ is the path loss exponent. Additionally, ${w}_{\text{j}}$ is zero-mean complex Gaussian noise with unity variance, ${x}$ is the unity energy transmitted symbol, while ${\tilde{x}}$ is the unity energy symbol forwarded by the relay. Let us remark that use a different codebook and $x$ and ${\tilde{x}}$ may be not identical.
\subsubsection{FD Relaying}
\begin{figure}[!t]
\centering
\includegraphics[width=.9\columnwidth]{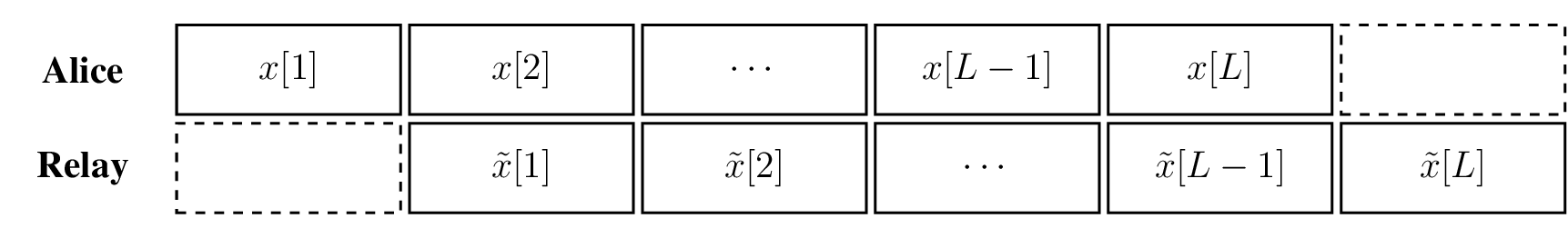}
\vspace{-2mm}
\caption{Alice and relay block transmission. Notice that the frame is divided into $L$ blocks and the relay's transmission is delayed by one block ($\tau=1$).}
\label{fig:block_relay_Tx}
\vspace{-4mm}
\end{figure}

The message is divided into $L$ blocks, as shown in Fig.~\ref{fig:block_relay_Tx}. Additionally, due to the inherent characteristics of the encoding/decoding scheme, ${\tilde{x}}$ is delayed compared to $x$, so that ${\tilde{x}}[l] = x[l - \tau]$, where $1 \leq l \leq L$ and $\tau$ represents the processing and block delay of $\tau \geq 1$ blocks, which we assume hereafter to be $\tau=1$. As pointed out in \cite{ART:RIIHONEN-TWC2011} this delay is large enough to guarantee that the received signals are uncorrelated, and therefore can be jointly decoded. A first analysis on such decoding schemes for practical FD relaying is done in \cite{AlvesETT2013}, which is later extended in \cite{Khafagy2013}, where the authors generalize the backward decoding scheme for any delay and number of blocks and show that performance is not affected for large $L$. Hereafter we assume regular encoding and backward decoding \cite{Kramer2005,LifengIT2008}\footnote{Please refer to \cite{Kramer2005} for details on DF regular encoding and backward decoding, and to \cite[Th. 2]{LifengIT2008} for an analysis on the relay-eavesdropper channel.}. 

Further, we consider all channels ($h_\text{ij}$) as quasi-static independent and identically distributed (i.i.d.) according to
composite Nakagami-$m$ Log-Normal distribution. The relay suffers from self-interference, which we model as a composite fading channel denoted by $h_{\text{RR}}$, and $\delta$ represents the overall self-interference attenuation factor. Additionally, since all RVs are i.i.d. the instantaneous SNRs of the legitimate links are LN distributed as $\Gamma_{\text{ij}} \thicksim \mathsf{LN}\left(\mu_{\text{ij}}, \sigma_{\text{ij}}\right)$ \cite{ART:HO-ACM95}. The SINR at the relay is $\Gamma_\text{R} = \frac{\Gamma_{\text{AR}}}{\Gamma_{\text{RR}}}$, which is also a LN RV defined as $\Gamma_\text{R} \thicksim \mathsf{LN}(\mu_\text{R} \,,\, \sigma_\text{R})$, where $\mu_\text{R}=\mu_\text{AR}-\mu_\text{RR}$ and $\sigma_\text{R}= (\sigma_\text{AR}^2+\sigma_\text{RR}^2)^\frac{1}{2}$ \cite{papoulis}.
\review{The first two cumulants of $\Gamma_\text{ij}$ can be readily attained through Definition~\ref{def:additivity} as $\kappa_{1,{\Gamma_\text{ij}} }$ and $\kappa_{2,{\Gamma_\text{ij}}}$.} 
With that, and relying on the additive property \cite{papoulis}, we obtain the cumulants of the overall SNR at Bob, $\Gamma_\text{B} = \Gamma_\text{AB}+\Gamma_\text{RB} $, and then the sum of two LN RVs can be simply written as a single LN RV as $\Gamma_\text{B} \thicksim \mathsf{LN}(\mu_\text{B}, \sigma_\text{B})$, \review{where $\mu_\text{B}=\ln{\left( \kappa_{1,\Gamma_\text{B}}^2\right)}-\tfrac{1}{2}\ln({\kappa_{1,\Gamma_\text{B}}^2 + \kappa_{2,\Gamma_\text{B}}})$ and $\sigma_\text{B}^2=\ln{\left( \kappa_{1,\Gamma_\text{B}}^2 + \kappa_{2,\Gamma_\text{B}} \right)}-\ln(\kappa_{1,\Gamma_\text{B}}^2)$
}.

\subsection{Eavesdropper Channel}
We assume that Eve is equipped with $N_\text{E}$ antennas and applies MRC to the received signals. Thus, the $N_\text{E}\times 1$ received signal is
\begin{align}
\label{eq:yAE}
{\bf y}_{\text{E}} &= \sqrt{P_\text{A} d_{\text{AE}}^{-\nu}} \, {\bf h}_{\text{AE}} \cdot {\bf x} +\,\sqrt{ P_\text{R}\,d_{\text{RE}}^{-\nu} } \, {\bf h}_{\text{RE}} \cdot {\bf \tilde{x}} + {\bf w}_{\text{E}},
\end{align}
where ${\bf h}_{\text{iE}}$, $\text{i} \in \{\text{A} , \text{R} \}$, denotes the channel coefficients vectors ($N_\text{E} \times 1$) at the eavesdropper. Additionally, ${\bf w}_{\text{E}}$ is zero-mean complex Gaussian noise $N_\text{E}\times 1$ vector with unity variance. 

\review{
From Definition~\ref{def:additivity} we obtain the cumulants of $\gamma_{\text{iE}_k}$. Moreover, once all RVs are i.i.d., we the cumulants of $\Gamma_\text{iE}=\sum_{k} \gamma_{\text{iE}_k}$ are simply $N_\text{E}\, {\kappa}_{n,\Gamma_\text{iE}}$, where $i \in \{\text{A, R}\}$ and $1\leq k \leq N_\text{E}$.}
Then, since $\Gamma_\text{AE}$ and $\Gamma_\text{RE}$ are independent we resort again to  Definition~\ref{def:additivity} and define the cumulants of $\Gamma_\text{E} = \Gamma_\text{AE} +\Gamma_\text{RE}$ as \review{ $\kappa_{n} = N_\text{E} ({\kappa}_{n,\Gamma_\text{AE}} + {\kappa}_{n,\Gamma_\text{RE}})$, which allows us to write $\Gamma_\text{E} \thicksim \mathrm{LN}(\mu_\text{E} \,,\, \sigma_\text{E})$, whose parameters are $\mu_\text{E}= \ln{\left(\kappa_{1,\Gamma_\text{E}}^2\right)}-\tfrac{1}{2} \ln({\kappa_{1,\Gamma_\text{E}}^2 + \kappa_{2,\Gamma_\text{E}}})$ and $\sigma_\text{E}^2 = {\ln{\left( \kappa_{1,\Gamma_\text{E}}^2 + \kappa_{2,\Gamma_\text{E}} \right)}-\ln(\kappa_{1,\Gamma_\text{E}}^2)}$.
}
%

\section{Scenario 1 (Sc1): Average Secrecy Rate}\label{sect:erg_cap}
First, let us suppose that Eve is part of the network and Alice is able to acquire the CSI from it. For instance, this scenario may represent Alice wanting to communicate privately to a certain user (\emph{i.e.}, Bob) without being overheard by other legitimate receivers. Thus, Alice can adapt its transmission rate accordingly in order to achieve perfect secrecy. In such a scenario, the average secrecy capacity is an insightful metric once it quantifies the average secrecy rate \cite{BlochBarros2008}.
\begin{proposition}
The achievable rates of the DF protocol for the relay-eavesdropper channel in the presence of self-interference is ${\cal R}_s = \left[ \log_2 \left( 1 + \Gamma_\text{FD} \right) - \log_2 \left( 1 + \Gamma_\text{E} \right) \right]^+ \,$, where $[x]^+ \triangleq \max\left\lbrace x , 0 \right\rbrace$.
\end{proposition}
\begin{IEEEproof}
Based on the average rate derivation for traditional relaying in \cite{Beaulieu2009}, plus given the cooperative secrecy rates from \cite{LifengIT2008}, and given that all RVs are independent we attain
\begin{equation}
\begin{split}
\label{eq:rs_avg2}
{\cal R}_s &= \left[{\cal R}_{\text{FD}} - {\cal R}_{\text{E}} \right]^+, \\
&= \left[ \log_2 \left( 1 + \min\left\lbrace \Gamma_\text{R} \,,\, \Gamma_\text{B}  \right\rbrace \right) - \log_2 \left( 1 + \Gamma_\text{E}  \right) \right]^+ \,.
\end{split}
\end{equation}
Let us introduce a new RV defined as $\Gamma_\text{FD} = \min\left\lbrace  \Gamma_\text{R} \,,\, \Gamma_\text{B} \right\rbrace$, whose CDF is introduced in Appendix~\ref{app:PDFCDFGammaFD}, then plugging it into \eqref{eq:rs_avg2} we conclude the proof. 
\end{IEEEproof}
\begin{theorem}\label{lm:sec_rate}
Assuming the non-negativity of the secrecy rate and the independence of RVs, the average secrecy rate is 
\begin{equation}
\begin{split}
\overline{\cal R}_s &= \int_{0}^{\infty} \int_{0}^{\infty} {\cal R}_s \, f_{\Gamma_\text{FD}}(\gamma_\text{FD}) \, f_{\Gamma_\text{E}}(\gamma_\text{E}) \, \mathrm{d}\gamma_\text{FD} \, \mathrm{d}\gamma_\text{E}, \\
\label{eq:eqAvgRateGeral}
&= \frac{1}{\ln 2} \int_{0}^{\infty} \frac{F_{\Gamma_\text{E}}(\gamma_\text{E}) }{1+\gamma_\text{E} } \left[1-  F_{\Gamma_\text{FD}}(\gamma_\text{E}) \right] \mathrm{d}\gamma_\text{E}.
\end{split}
\end{equation}
\end{theorem}
\begin{IEEEproof}
See \cite{LifengTIFS2014}. 
\end{IEEEproof}

Next, since the CDFs of $\Gamma_\text{FD}$ and $\Gamma_\text{E}$ are known and with help of Theorem~\ref{lm:sec_rate} we attain the average secrecy rate (bits/s/Hz) as 
\begin{align}\label{eq:integralRs}
\overline{\cal R}_s &= \frac{1}{\ln 2} \int_{0}^\infty \frac{\erfc{ \eta_\text{E} } \, \erfc{ \eta_\text{B} } \,
\erfc{ \eta_\text{R} } }{8(1+z)} \mathsf{d}z \,,
\end{align}
where $\eta_\text{E}=\frac{\mu_\text{E}-\ln(z)}{\sqrt{2}\,\sigma_\text{E}}$, $\eta_\text{B}=\frac{-\mu_\text{B}+\ln(z) }{\sqrt{2}\,\sigma_\text{\text{E}}}$ and $\eta_\text{R} = \frac{-\mu_\text{R}+\ln(z)}{\sqrt{2}\,\sigma_\text{R}} $. Nevertheless, \eqref{eq:integralRs} does not have a closed form solution, therefore we resort to Gauss-Laguerre quadrature \cite[Ch. 25.4]{ABRAM65}.
\begin{theorem} Under the assumption of perfect CSI  and composite fading, the average secrecy rate in bits/s/Hz is 
\begin{align}\label{eq:avgSecRate}
\overline{\cal R}_s \simeq \frac{1}{\ln 2} \sum_{k=1}^{K} \frac{\omega_k^{\text{L}} \, e^{\chi_k^{\text{L}}}}{8} \erfc{\eta_\text{E}^\prime} \erfc{\eta_\text{B}^\prime} \erfc{\eta_\text{B}^\prime},
\end{align}
where $\eta_\text{i}^\prime$ with $\text{i} \in \{\text{B}\,,\text{E}\,,\text{R}\}$ is written as in $\eta_\text{i}$ but replacing $z$ by $e^{\chi_k^{\text{L}}}-1$, and $K$ is the order of the Laguerre polynomial. 
\label{th:avgSecRate}
\end{theorem}
\begin{IEEEproof}
Please see \cite{LifengTIFS2014} and notice that we resort to Gauss-Laguerre quadrature \cite[Ch. 25.4]{ABRAM65} to attain \eqref{eq:avgSecRate}, thus $\chi_k^{\text{L}}$ are the roots of the Laguerre polynomial while $\omega_k^{\text{L}} = \chi_k^{\text{L}} \,/\, \left( (K+1)\,\operatorname{L}_{K+1}(\chi_k^{\text{L}}) \right)^2$ are the weights of the Gauss-Laguerre quadrature \cite[Ch. 25.4]{ABRAM65}. The error can be analytically estimated as indicated in \cite[Ch. 25.4]{ABRAM65}.
\end{IEEEproof}

\section{Scenario 2 (Sc2): Secrecy Outage Probability}\label{sect:out_prob_SC2}
Now let us suppose that Alice has no knowledge of Eve's CSI except for channel statistics. Thus, in order to protect its transmission from a possible inimical attack, Alice communicates with Bob with a constant secrecy rate ${R}_s >0$, which yields a certain secrecy outage probability\footnote{In order to avoid confusion it is noteworthy that the secrecy capacity is denoted by the calligraphic letter $\cal C$ and the secrecy achievable rate by letter $\cal R$, while the attempted transmission rate is represented by the letter $R$.}. Note that Alice assumes that Eve' secrecy rate is ${\cal R}_\text{E} = {\cal R}_\text{FD} - {R}_s$, thus security is compromised as soon as ${\cal C}_\text{E}>{\cal R}_\text{E}$. Otherwise, perfect secrecy is assured. Then, secrecy outage probability is the appropriated metric to evaluate the performance of a quasi-static fading wiretap channel when the transmitter has no CSI and the receivers have CSI of their own channels only \cite{BlochBarros2008}. Next, we determine the secrecy outage probability as ${\cal O}_\text{P} = \Pr\left[{\cal R}_s < {R}_s\right]$. Thus, an outage event occurs whenever the instantaneous secrecy rate ${\cal R}_s$ falls below the target secrecy rate $R_s$ fixed by Alice.
\begin{theorem}
\label{lm:OUTAGE_SC2}
Assuming the non-negativity of the secrecy rate and that only CSI of the legitimate channel is available at the transmitter, we can define the secrecy outage probability in bits/s/Hz as
\begin{align}
\label{eq:outageSC2_1}
{\cal O}_\text{P} \simeq 1-\sum\limits_{k=1}^{K} \frac{\omega_{k}^{\text{H}} \erfc{ \frac{-\mu_\text{B}+\ln(\upsilon_\text{E}) }{ \sqrt{2}\,\sigma_\text{B} } } \erfc{ \frac{-\mu_\text{R}+\ln(\upsilon_\text{E}) }{ \sqrt{2}\,\sigma_\text{R} } } }{4\sqrt{\pi}} \, ,
\end{align}
where $\upsilon_\text{E} = 2^{R_s}\left( \exp\left(\mu_\text{E}+\sqrt{2} \sigma_\text{E} \,\chi_k^\text{H} \right) + 1 \right)-1$. 
\end{theorem}
\begin{IEEEproof}
Based on \cite{LifengIT2008,BlochBarros2008}, we can define the secrecy outage probability under composite fading channels as 
\begin{align} 
{\cal O}_{P}&=\! \Pr\left[\! {\cal R}_s \! <\! {R}_s\right]\! 
=\! \Pr\left[\! \log_2 \left(\! 1 \!+ \gamma_\text{FD} \right)\! - \log_2 \left(\! 1 \! + \gamma_\text{E} \right)\! < {R}_s \right] \nonumber \\
\label{eq:outSC2_integral}
&= \int\limits_0^\infty F_{\Gamma_\text{FD}}(2^{R_s}(1+z)-1) f_{\text{E}}(z) \mathsf{d}z.
\end{align}
Note that the CDF of $\Gamma_\text{FD}$ is given in Appendix~\ref{app:PDFCDFGammaFD}. However, \eqref{eq:outSC2_integral} does not have a closed-from solution. Therefore, we resort to semi-analytical solution based on Gauss-Hermite quadrature \cite[Ch. 25.4]{ABRAM65}. The weights of the Hermite polynomial are given as $\omega_k^\text{H} = \sqrt{\pi}\,2^{K-1}\,K! \,/\, \left( K^2 \, \operatorname{H}_{K-1}(\chi_k^\text{H}) \right)$, where $\chi_k^\text{H}$ are its roots of order $K$ \cite[Ch. 25.4]{ABRAM65}. 
\end{IEEEproof}

\section{Numerical Results and Discussions}\label{sect:num_res}
Alice, the relay and Bob are assumed to be on a straight line, the relay is at the center and the distance between Alice and Bob is $30$m, unless stated otherwise. Additionally, we assume a path loss exponent of $\nu=4$ as well as unitary bandwidth. We assume that all channels experience some LOS, and therefore undergo fading with $m=2$, while the shadowing standard deviation is $\sigma=10$dB\footnote{We assume that the polynomial order of $K=24$ since it presents great accuracy. We recall that the error can be analytically estimated which is another advantage of such quadrature methods~\cite{ABRAM65}.}. We consider equal power allocation thus total power is given as $P_\text{A}=P_\text{R}=P$. 

\begin{figure}[!t]
\centering
\includegraphics[width=1\columnwidth]{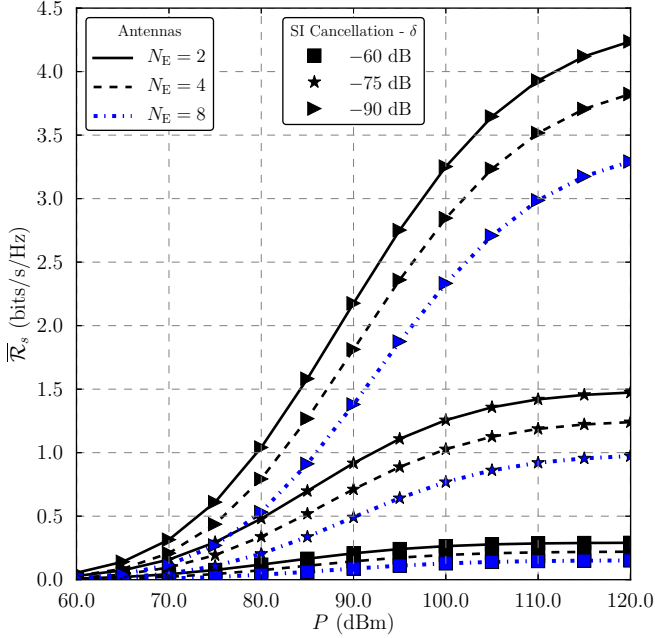}
\caption{Average secrecy rate $\overline{{\cal R}}_s$ as a function of the transmit power $P$ in dBm for different values of self-interference cancellation ($\delta$) and number of antennas $N_\text{E}$.}
\label{fig:avgRSxPsNE248}
\vspace{-4mm}
\end{figure}

Fig.~\ref{fig:avgRSxPsNE248} depicts the average secrecy rate (Sc1) as a function of transmit power $P$ in dBm for different values of self-interference cancellation for $N_\text{E}\in\{2,4,8\}$ antennas. Eve's average SNR is set to $\mu_\text{E}=.21$ and $\sigma_{E}=.76$, which means that $\mu_{\Omega_p}=-10$~dB and $\sigma_{\Omega_p}=5$~dB so that Eve is closer to Alice and the relay than Bob. Notice that the more sophisticated is the self-interference cancellation (lower values of $\delta$), higher is the achievable average secrecy rate. For instance, at high SNR, the average secrecy rate can be doubled with the reduction of the self-interference from -80dB to -90dB. Even though sophisticated interference mitigation schemes have been recently proposed,  such cancellation levels -- in the order of -90dB -- are still a challenging task to achieve \cite{DuarteFDTWC2012}. The average secrecy rate degrades as the number of antennas grows. Nevertheless, it is still possible to communicate with perfect secrecy. Notice that in terms of secrecy capacity the impact of relatively poor self-interference cancellation ($\delta > -75$~dB) is much worse than that of having a sophisticated eavesdropper (with several receive antennas applying maximum ratio combining).

\begin{figure}[!t]
\centering
\includegraphics[width=1\columnwidth]{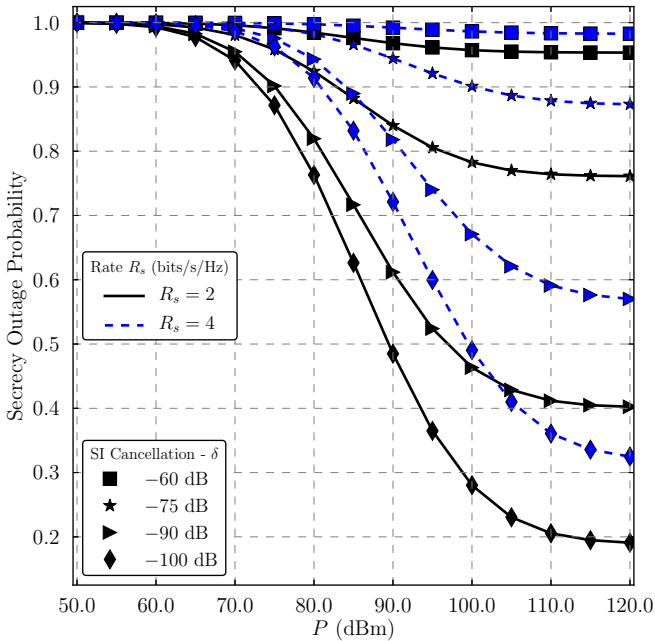}
\caption{Outage probability, $\Pr\left[{\cal R}_s < R_s \right]$, as a function of the transmit power $P$ in dBm with different values of self-interference cancellation ($\delta$) as well as target secrecy rate $R_s$.}
\label{fig:OUT_Power_RS}
\vspace{-4mm}
\end{figure}

Further, Fig.~\ref{fig:OUT_Power_RS} shows the secrecy outage probability (Sc2) as a function of the transmit power for  $R_s \in \{2,4\}$ bits/s/Hz. Secrecy outage probability considerably increases with better self-interference attenuation and cancellation at the relay (lower $\delta$). Notice also that as the target secrecy rate increases, the secrecy outage probability decreases since Alice imposes a more stringent secure rate requirement.
%
\section{Conclusions}\label{sect:conc}
We investigated the secrecy performance of FD relaying in the presence of a multiple antenna eavesdropper under composite fading. Two scenarios are considered based on the CSI availability at the transmitters: \textit{a)} full CSI is available from both intended and unintended receivers; \textit{b)} only the legitimate channels are known to Alice. Our results show that the self-interference at the relay considerably affects performance regardless of the scenario. However, even under strong self-interference and in the presence of a sophisticated multiple antenna eavesdropper, FD relaying is feasible.
%
\appendix
%
\label{app:PDFCDFGammaFD}
Let $\Gamma_\text{R}$ and $\Gamma_\text{B}$ be independent and LN distributed, thus CDF of $\Gamma_\text{FD} = \min\left\lbrace  \Gamma_\text{R} \,,\, \Gamma_\text{B} \right\rbrace$ can be readily attained through standard statistical methods \cite{papoulis}, such that the CDF is
\begin{equation}\label{eq:cdfGammaFD}
F_{\Gamma_\text{FD}}(z) 
=\! 1- \frac{1}{4} \erfc{\tfrac{-\mu_\text{R}+\ln(z)}{\sqrt{2}\,\sigma_{R}}}\erfc{\tfrac{-\mu_\text{B}+\ln(z)}{\sqrt{2}\,\sigma_{B}}}.
\end{equation}

\bibliographystyle{IEEEtran}

\end{document}